\documentclass[prx,twocolumn,amsmath,amssymb,superscriptaddress,showpacs]{revtex4-1}

\usepackage{graphicx}
\usepackage{dcolumn}
\usepackage{bm}

\usepackage{color}

\newcommand{\bra}{\langle}
\newcommand{\ket}{\rangle}

\renewcommand{\vec}[1]{\boldsymbol{\mathbf{#1}}}

\begin{document}

\title{Effect of spin-orbit coupling on the high harmonics from the topological Dirac semimetal Na$_3$Bi}

\author{Nicolas Tancogne-Dejean}
  \email{nicolas.tancogne-dejean@mpsd.mpg.de}
 \affiliation{Max Planck Institute for the Structure and Dynamics of Matter and Center for Free-Electron Laser Science, Luruper Chaussee 149, 22761 Hamburg, Germany}
 \affiliation{European Theoretical Spectroscopy Facility (ETSF)}
 
\author{Florian G. Eich}
 \affiliation{HQS Quantum Simulations GmbH, Haid-und-Neu-Stra{\ss}e 7, D-76131 Karlsruhe, Germany}

 \author{Angel Rubio}
  \email{angel.rubio@mpsd.mpg.de}
\affiliation{Max Planck Institute for the Structure and Dynamics of Matter and Center for Free-Electron Laser Science, Luruper Chaussee 149, 22761 Hamburg, Germany}
 \affiliation{European Theoretical Spectroscopy Facility (ETSF)}
 \affiliation{Nano-Bio Spectroscopy Group, Universidad del Pa\'is Vasco, 20018 San Sebasti\'an, Spain }
\affiliation{Center for Computational Quantum Physics (CCQ), The Flatiron Institute, 162 Fifth Avenue, New York NY 10010}

\begin{abstract}
In this work, we performed extensive first-principles simulations of high-harmonic generation in the topological Diract semimetal Na$_3$Bi using a time-dependent density functional theory framework, focusing on the effect of spin-orbit coupling (SOC) on the harmonic response. We also derived a general analytical model describing the microscopic mechanism of strong-field dynamics in presence of spin-orbit coupling, starting from a locally $U(1)\times SU(2)$ gauge-invariant Hamiltonian. Our results reveal that SOC: (i) affects the strong-field ionization by modifying the bandstructure of Na$_3$Bi, (ii) modifies the electron velocity, making each spin channel to react differently to the pump laser field, (iii) changes the emission timing of the emitted harmonics. Moreover, we show that the SOC affects the harmonic emission by directly coupling the charge current to the spin currents, paving the way to the high-harmonic spectroscopy of spin currents in solids.
\end{abstract}

\maketitle
\section{Introduction}

The swift development of strong-field attoscience in solids has recently allowed for the development of many novel techniques, tailored toward controlling electron dynamics in solids on unprecedented timescales.
Among the possible applications of controlled strong-field dynamics in solids, one finds the possibility to perform high-harmonic spectroscopy of various fundamental phenomena with unprecedented timescales, such as  bandstructure dynamics~\cite{PhysRevLett.115.193603,Lanin:17,PhysRevLett.124.157403,Borsch1204,PhysRevLett.118.087403,tancogne2017ellipticity}, dynamical correlation effects~\cite{PhysRevLett.121.097402,silva2018high,PhysRevLett.124.157404}, or the study of structural and topological phase transitions~\cite{PhysRevB.102.134115,silva2019topological}. Given the wealth of physical phenomena occurring in solids, much more exciting results are expected to emerge in the coming years.\\
Recently, topological effects, and the related Berry curvature effects, have attracted a lot of attention in the context of strong-field dynamics\cite{PhysRevB.102.174314,jimenez2020lightwave,PhysRevLett.120.177401,PhysRevB.100.125144,bai2020high}. In particular, several studies investigated how high-harmonic generation (HHG) is affected by a topological phase transition using the Haldane model~\cite{PhysRevB.102.134115,silva2019topological,PhysRevA.102.053112} and related Haldane nanoribons models~\cite{PhysRevA.102.043105}.
In order to demonstrate the possibility to probe topological phase transitions using high-harmonic generation in a experiment, one needs to find a material that can host these different phases, and for which one can, by tuning an external parameter, reach different parts of its phase diagram. In this respect, Na$_3$Bi was shown to be a very promising material, as its phase diagram is extremely rich and displays different topologically non-trivial phases, such as topological Dirac and Weyl semimetal phases, topological and trivial insulating phases, or a phase with non-trivial Fermi surface states with non-zero topological charges~\cite{PhysRevB.85.195320}. 
Without breaking any system symmetry, Na$_3$Bi is a three-dimensional (3D) topological Dirac semimetals\cite{liu2014discovery} where two overlapping Weyl fermions form a 3D Dirac point. 
It was shown that a compression of 1\% along the $y$ axis, which breaks inversion symmetry, can turn Na$_3$Bi into a topological insulator. Moreover, \textit{ab initio} calculations for Na$_3$Bi driven by a circularly-polarized light showed that circularly-polarized light, by breaking time-reversal symmetry, induces a Floquet-Weyl semimetal where the two Weyl points are separated in momentum space~\cite{hubener2017creating}. Importantly, because in this case the crystal symmetries of Na$_3$Bi are still preserved, these Weyl points remain topologically protected\cite{PhysRevB.85.195320}.
The capability to selectively break some of the symmetries of Na$_3$Bi either using pressure, or ultrafast circularly-polarized light pulses, allows to explore its phase diagram in depth. However, before exploring the complex phase diagram of Na$_3$Bi, one needs to get a deeper understanding of the strong-field response in Na$_3$Bi in its pristine phase. This is the main scope of the present work.\\
At the hearth of most of topological properties and Berry-curvature-related phenomena lies the spin-orbit coupling (SOC) interaction.
While SOC does not affect strongly HHG in atomic systems, mostly affecting the harmonic yield\cite{pabst2014spin}, it is known that the SOC can strongly modify the bandstructure of bulk materials, and is responsible for various phenomena such as spin-Hall currents~\cite{RevModPhys.87.1213}, the anomalous Hall effect~\cite{RevModPhys.82.1539}, and spintronics~\cite{RevModPhys.76.323}, to cite a few.
Understanding how the SOC affects the strong-field dynamics in solids is therefore of paramount importance in order to later understand related phenomena such as topology. This is also why Na$_3$Bi is a very attractive choice, as it allows us to investigate SOC effects, without any Berry-curvature-related phenomena.\\
It is clear that SOC can modify the bandstructure of materials, and can, for instance, lead to band-inversion and thus induce topological insulators. SOC can also split energy bands and lift spin degeneracy, and renormalize electronic bands. Besides these modifications of the bandstructure, one might wonder if and how the SOC affects the dynamics of the excited carriers within these modified bands.
Already, by considering the lowest relativistic correction to the Pauli Hamiltonian containing the spin-orbit interaction $H_{\mathrm{SO}} = \frac{\hbar e}{4m^2c^2}\vec{p}\ldotp(\vec{\sigma}\times\vec{\nabla}V)$, where $V$ is the potential due to the ions acting on the electrons, $\vec{p}$ their momentum, one finds that the velocity operator is modified such that $\hat{\vec{v}} = -\frac{i}{\hbar}[\hat{\vec{r}},\hat{H}] = \frac{1}{m}[\vec{p}+\frac{e}{c}\mathcal{A}]$, where $\mathcal{A} = \frac{\hbar}{4mc}\vec{\sigma}\times\vec{\nabla}V$ plays the role of a SU(2) gauge vector potential\cite{PhysRevLett.95.187203}. This can be seen as leading to an effective magnetic field $\mathcal{B}=\vec{\nabla}\times\mathcal{A} = \hbar [\vec{\sigma}\ldotp(\nabla^2V) - (\vec{\sigma}\ldotp\vec{\nabla})\vec{\nabla}V]/(4mc)$. How this change in the electrons' velocity reflects in the strong-field response of solids is one of the main focus of this work. As we will see later, this change in the electrons' velocity leads to a coupling between the charge current and the microscopic fluctuations of the spin current.

The relation between spin-current and nonlinear response of bulk material has already been partly investigated in the context of perturbative nonlinear optics.
It was shown that a pure spin-current can induce a nonlinear response in GaAs in the form of a non-zero perturbative second-harmonic generation along specific directions ~\cite{PhysRevLett.104.256601,werake2010observation}. It is therefore interesting to ask if HHG in solids could be used to measure spin currents.
In Ref.~\cite{zhang2018generating}, authors investigated theoretically the optical and spin harmonics emitted from iron monolayers, showing that the spin current displays a similar harmonic structure as the HHG obtained from the charge current. Whereas they showed that the spin current also exhibit non-perturbative harmonics that depends on the specific symmetries of their system, a discussion on  how the HHG is directly influenced  by the spin current remains elusive.
To answer this question, we derived here the exact equation of motion of the physical charge current starting from a locally U(1)$\times$SU(2) gauge invariant Hamiltonian,\cite{RevModPhys.65.733,PittalisEich:17} including spin-orbit coupling.


%
%

The outline of this paper is as follows. First, in Sec.~\ref{sec:method}, we give the details of the \textit{ab initio} method used to model the electron dynamics in Na$_3$Bi and we discuss some of its electronic properties. Then, in Sec.~\ref{sec:results}, we study the strong-field electron dynamics in Na$_3$Bi, focusing on the effect of spin-orbit coupling on its strong-field response. In Sec.~\ref{sec:EOM}, we present the derivation of the equation of motion of the conserved charge current for a U(1)$\times$SU(2) locally gauge-invariant Hamiltonian, including necessary relativistic corrections. In particular, we obtain, in the lowest order in $1/c$ a new formula for HHG in presence of spin-orbit coupling for non-magnetic materials. Finally, we draw our conclusions in Sec.~\ref{sec:conclusions}.

\section{Method}
\label{sec:method}

%
Calculations were performed using an in-plane lattice parameter of $a=$5.448~\AA\ and an out-of-plane lattice parameter $c=$9.655\AA\, taking as the structure the one of Ref.~\cite{PhysRevB.85.195320}, a real-space spacing of $\Delta r=0.36$ Bohr, and a $28\times28\times15$ $\vec{k}$-point grid to sample the Brillouin zone. 
We employ norm-conserving fully relativistic Hartwigsen-Goedecker-Hutter (HGH) pseudo-potentials, and we included spin-orbit coupling unless stated differently. In all the calculations, the ions are kept fixed and the coupled electron-ion dynamics is not considered in what follows.
For such few-cycle driver pulses, the HHG spectra from solids have been shown to be quite insensitive to the carrier-envelope phase (CEP), which is therefore taken to be zero here.
We consider a laser pulse of 25 fs duration (FWHM), with a sin-square envelope for the vector potential. The carrier wavelength $\lambda$ is 2100\,nm, corresponding to a carrier photon energy of 0.59\,eV and the intensity of the laser field is taken to be $I=4\times10^{11}$W.cm$^{-2}$ in matter.
The time-dependent wavefunctions and current are computed by propagating Kohn-Sham equations within TDDFT, as provided by the Octopus package.~\cite{Octopus_paper_2019}
The time-dependent Kohn-Sham equation within the adiabatic approximation reads 
\begin{eqnarray}
 i\hbar\frac{\partial}{\partial t}|\psi_{n,\mathbf{k}}(t)\rangle = \Big[\frac{(\hat{\mathbf{p}}+e\mathbf{A}(t)/c)}{2m} + \hat{v}_{\mathrm{ext}} + \hat{v}_{\mathrm{H}}[n(\mathbf{r},t)]\nonumber\\
 + \hat{v}_{\mathrm{xc}}[n(\mathbf{r},t)] + \hat{v}_{\mathrm{NL}}\Big]|\psi_{n,\mathbf{k}}(t)\rangle,
\end{eqnarray}
where  $|\psi_{n,\mathbf{k}}\rangle$ is a Pauli spinor representing the Bloch state with band index $n$, at the point $\mathbf{k}$ in the Brillouin zone, $\hat{v}_{\mathrm{ext}}$ is the electron-ion potential, $\mathbf{A}(t)$ is the external vector potential describing the laser field,  $\hat{v}_{\mathrm{H}}$ is the Hartree potential, $\hat{v}_{\mathrm{xc}}$ is the exchange-correlation potential, and $\hat{v}_{\mathrm{NL}}$ is the non-local pseudopotential, also describing the spin-orbit coupling.
In all our calculations, we employed the adiabatic local density approximation functional for describing the exchange-correlation potential.\\
From the time-evolved wavefunctions, we computed the total electronic current $\vec{j}(\vec{r},t)$,
\begin{equation}
 \vec{j}(\vec{r},t) = \frac{1}{2m} \sum_n \Re \Big[ \psi_n^*(\vec{r}) (\hat{\mathbf{p}}+e\mathbf{A}(t)/c) \psi_n(\vec{r}) \Big]\,,
\end{equation}

and from it, the HHG spectrum is directly obtained as
\begin{equation}
 \mathrm{HHG}(\omega) = \left|\mathrm{FT}\Big(\frac{\partial}{\partial t}\int d\vec{r}\, \vec{j}(\vec{r},t)\Big)\right|^2,
\end{equation}
where $\mathrm{FT}$ denotes the Fourier transform.\\
The total number of electron excited to the conduction bands ($n_{\mathrm{ex}}(t)$) can be obtained by projecting the time-evolved wavefunctions ($|\psi_n(t)\rangle$) on the basis of the ground-state wavefunctions ($|\psi_{n'}^{\mathrm{GS}}\rangle$) 
\begin{equation}
 n_{\mathrm{ex}}(t) = N_e - \frac{1}{N_\mathbf{k}}\sum_{n,n'}^{\mathrm{occ.}}\sum_{\mathbf{k}}^{\mathrm{BZ}} |\langle \psi_{n,\mathbf{k}}(t) | \psi_{n',\mathbf{k}}^{\mathrm{GS}} \rangle|^2,
\end{equation}
where $N_e$ is the total number of electrons in the system, and $N_{\mathbf{k}}$ is the total number of  $\mathbf{k}$-points used to sample the BZ. The sum over the band indices $n$ and $n'$ run over all occupied states.

%
%
%

%
\section{Results}
\label{sec:results}
%

\begin{figure}[ht]
  \begin{center}
    \includegraphics[width=0.95\columnwidth]{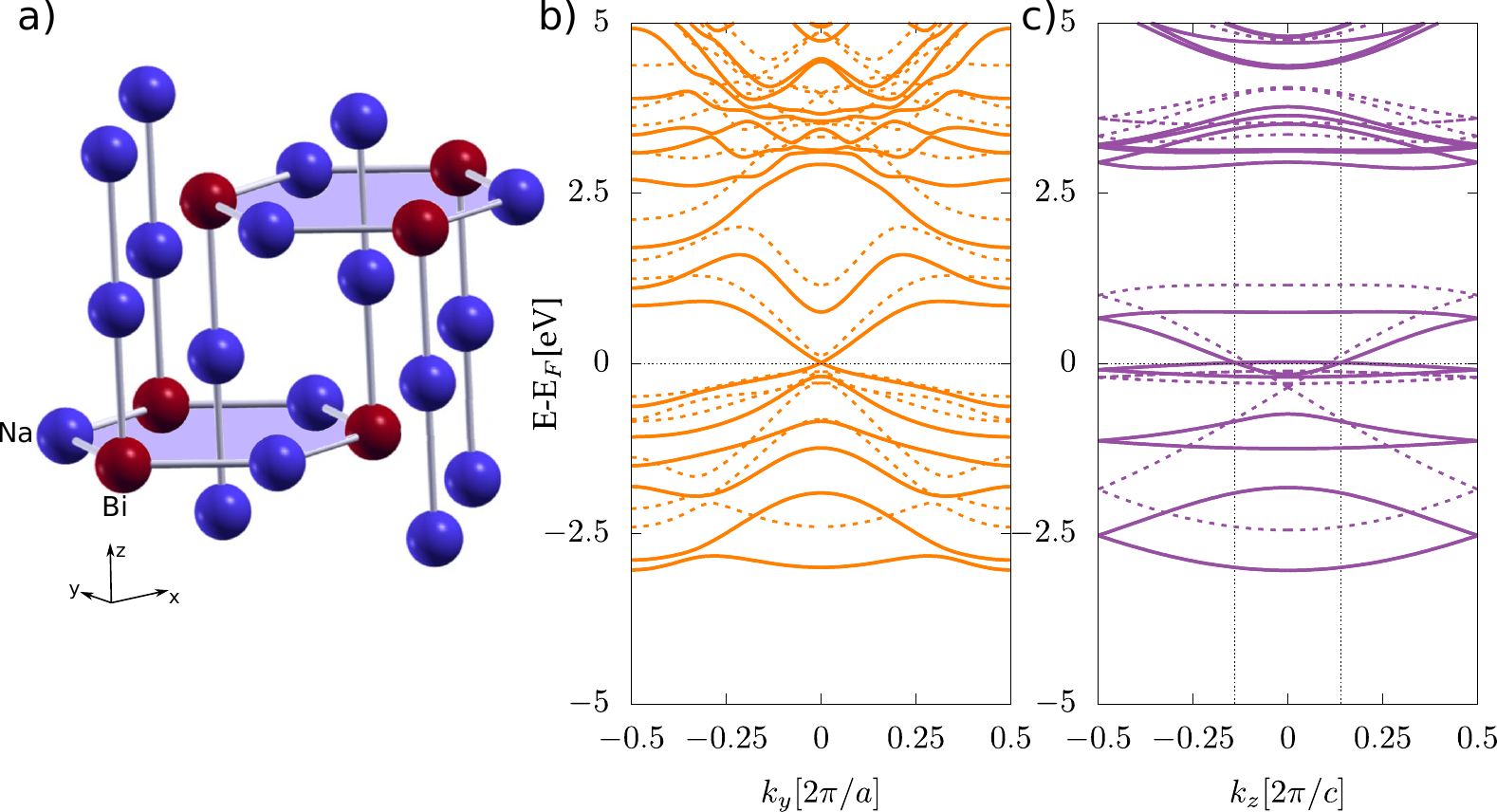}
    \caption{\label{fig:NaBi} a) Structure of bulk Na$_3$Bi. b) Band dispersion of Na$_3$Bi along the $k_y$ direction, for $k_x=0$ and $k_z=0.14\pi/c$, where $c$ is the out-of-plane lattice parameter. c) Band dispersion of Na$_3$Bi along the $k_z$ direction, for $k_x=k_z=0$. The vertical lines indicate the position of the two Weyl points. The dashed lines represent the bandstructure without SOC. }
  \end{center}
\end{figure}

Using our \textit{ab initio} TDDFT framework, we study HHG in bulk Dirac semimetals, taking Na$_3$Bi as a prototypical material. Its crystalline structure is shown in Fig.~\ref{fig:NaBi} a).
A key feature to describe the electronic structure of Na$_3$Bi is the SOC. The effect of SOC on the bandstructure of Na$_3$Bi is also shown in Fig.~\ref{fig:NaBi}. Without it, Na$_3$Bi is metallic, but does not display Dirac points at the Fermi energy, whereas SOC leads to the appearance of two Dirac cones at the Fermi energy, located at $(0,0,\pm 0.28\pi/c)$, in good agreement with the measured position of the Dirac cones in pristine Na$_3$Bi\cite{liu2014discovery}.\\

\subsection{HHG spectra}

\begin{figure}[ht]
  \begin{center}
    \includegraphics[width=0.9\columnwidth]{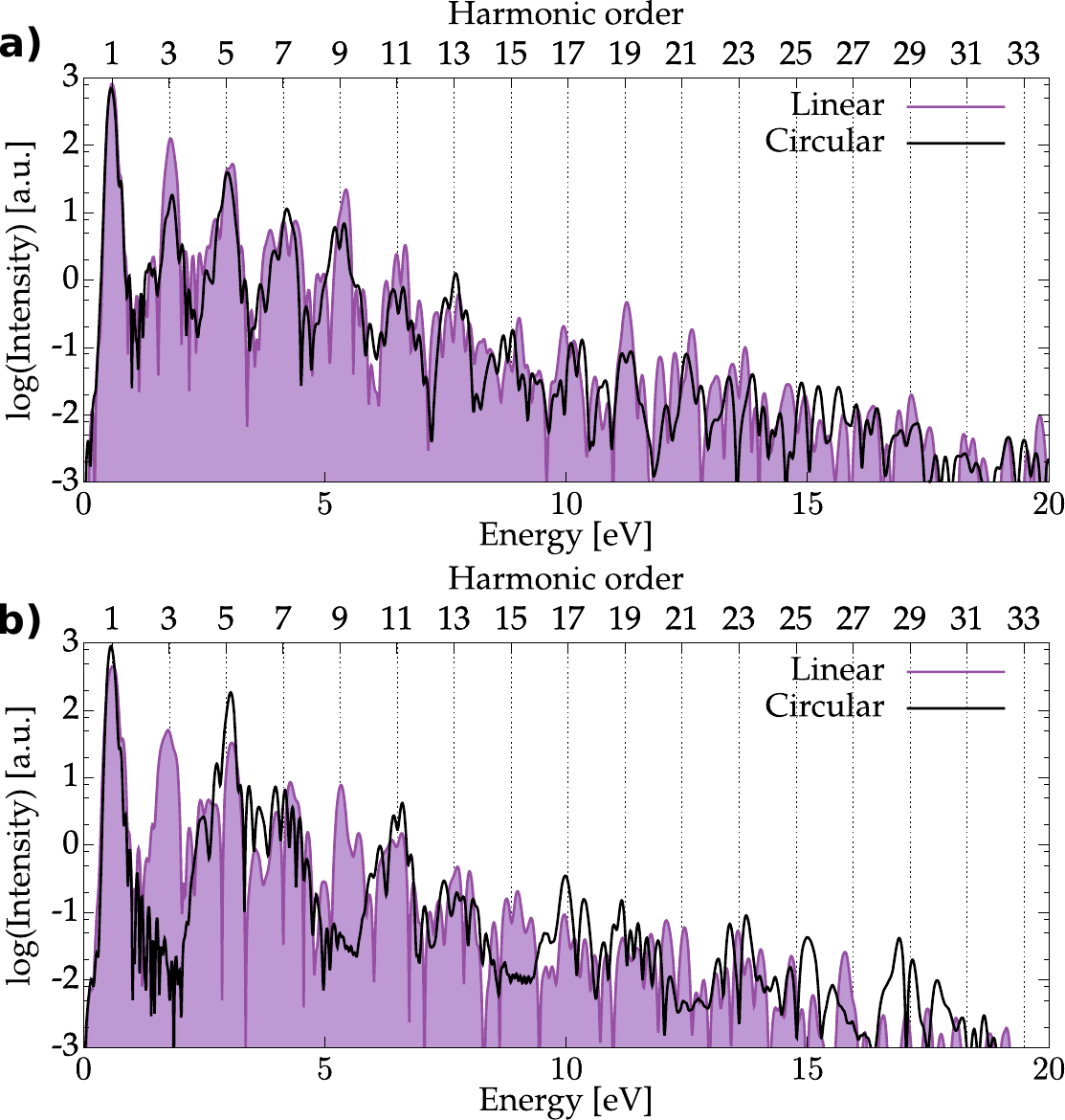}
    \caption{\label{fig:circular} Harmonic spectra obtained for linearly-polarized driving field (black curves) and for circularly-polarized driving field (red lines) for a) the $y-z$ polarization plane, and b) the $x-y$ polarization plane.}
  \end{center}
\end{figure}

We start by analyzing the harmonic response of the Na$_3$Bi when excited by linearly-polarized and circularly-polarized laser fields, with both $y-z$ and $x-y$ polarization planes, as shown in Fig.~\ref{fig:circular}. \\
Clear odd-structure harmonics are obtained in both cases, and few points can be noticed directly: i) the intensity of the harmonics emitted from a circularly-polarized driver is not really much smaller than the one obtained for a linearly-polarized driver. In fact, for a laser polarized in the $x-y$ plane (Fig.~\ref{fig:circular}b)) the harmonic emission is even stronger for the circularly-polarized driver. ii) the two polarization planes lead to qualitatively different results. Indeed for the $y-z$ plane, all odd harmonics are obtained for a circularly-polarized driver whereas for the $x-y$ case, we observe two consecutive odd harmonics over three, the last one been suppressed, which means we observe harmonics 5, 7 but not 9, and so on.
While this result looks surprising at first glance, it is only the result of selection rules~\cite{PhysRevB.3.4025} for different symmetries of the different polarization planes, as already discussed and confirmed experimentally by some of us ~\cite{tancogne2017ellipticity, klemke2019polarization} and others~\cite{saito2017observation}. Indeed, in the case of the $x-y$ polarization plane, the six-fold lattice symmetry leads to the suppression of the third harmonics, whereas the fifth and the seventh are not forbidden~\cite{PhysRevB.3.4025}, as observed for instance in quartz~\cite{klemke2019polarization}. In the case of the $y-z$ polarization, the four-fold symmetry leads to odd harmonics with alternating ellipticity, as observed in silicon or bulk MgO~\cite{tancogne2017ellipticity, klemke2019polarization}.

\subsection{Electron dynamics and SOC}

\begin{figure}[ht]
  \begin{center}
    \includegraphics[width=0.9\columnwidth]{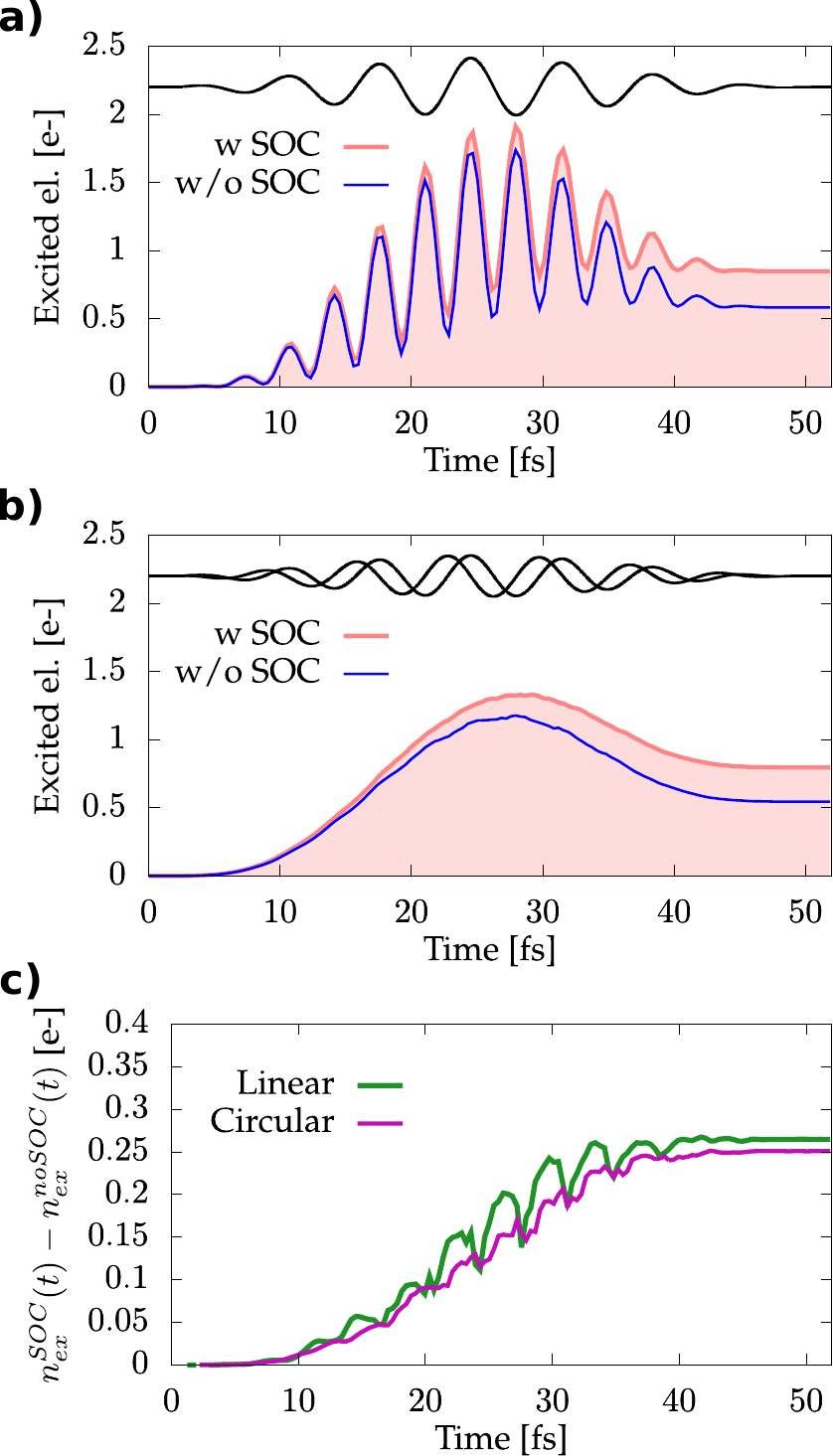}
    \caption{\label{fig:nex} Number of excited electrons for a) a linearly-polarized laser field along the $y$ axis, and b) a LHS circularly-polarized laser pulse with a major axis along the $y$ axis (middle panel), for an intensity in matter of $I_0=4\times10^{11}$W.cm$^{-2}$. In each panel, the blue curve corresponds to the result without SOC and the shaded area corresponds to the result including SOC. c) shows the difference in the number of excited electrons induced by SOC, for the linearly polarized and for the circularly polarized cases. In a) and b), the time evolution of the vector potential is shown in black.}
  \end{center}
\end{figure}

We now turn our attention to the number of electron excited by the laser pulse.
In an ionization process, the bandgap plays a key role in the excitation, and the ionization rate depends exponentially on it. We therefore expect that the number of excited electrons strongly depends on the details of the solid bandstructure in the case of interband ionization (Zener tunneling).
In order to reveal the role of SOC on the ionization, we computed the number of excited carriers in Na$_3$Bi for a linearly-polarized laser, as well as a left-hand-side (LHS) circularly polarized laser field, as shown in Fig.~\ref{fig:nex} for the same intensity.
This plot shows clearly that more electrons are excited to the conduction bands when the SOC is included, and that this does not really depend on the polarization state of the light (see bottom panel).
In the case of the linearly polarized case, we observe clear virtual excitations at twice the frequency of the laser field, whereas in the circularly-polarized case, these oscillations are not present. This can be understood as follow: in the circularly-polarized case, electrons are ionized twice more often than in the linearly polarized case during the laser pulse, every time each components of the field reaches an extrema. Moreover, because the field strength along each direction is reduced by a factor $1/\sqrt(2)$ for a circularly-polarized laser compared to the linearly-polarized case, ionization is roughly reduced by the same amount. This is what we expect from single-photon ionization through interband excitation. After the end of the pulse, the two pulses have still excited similar amount of electrons.
However, the contribution from the SOC, shown in Fig.~\ref{fig:nex}c) does not show a dependence on the polarization state nor the strength of the driver during the laser pulse, as it is almost identical for both linearly and circularly-polarized laser pulses, whose field strength differ by  a factor $1/\sqrt(2)$. We interpret this as originating not from interband transitions, but from an intraband acceleration of carriers, through the Dirac points that appear when SOC is included.


\subsection{Effect of SOC on the charge and spin currents}

\begin{figure}[ht]
  \begin{center}
    \includegraphics[width=0.9\columnwidth]{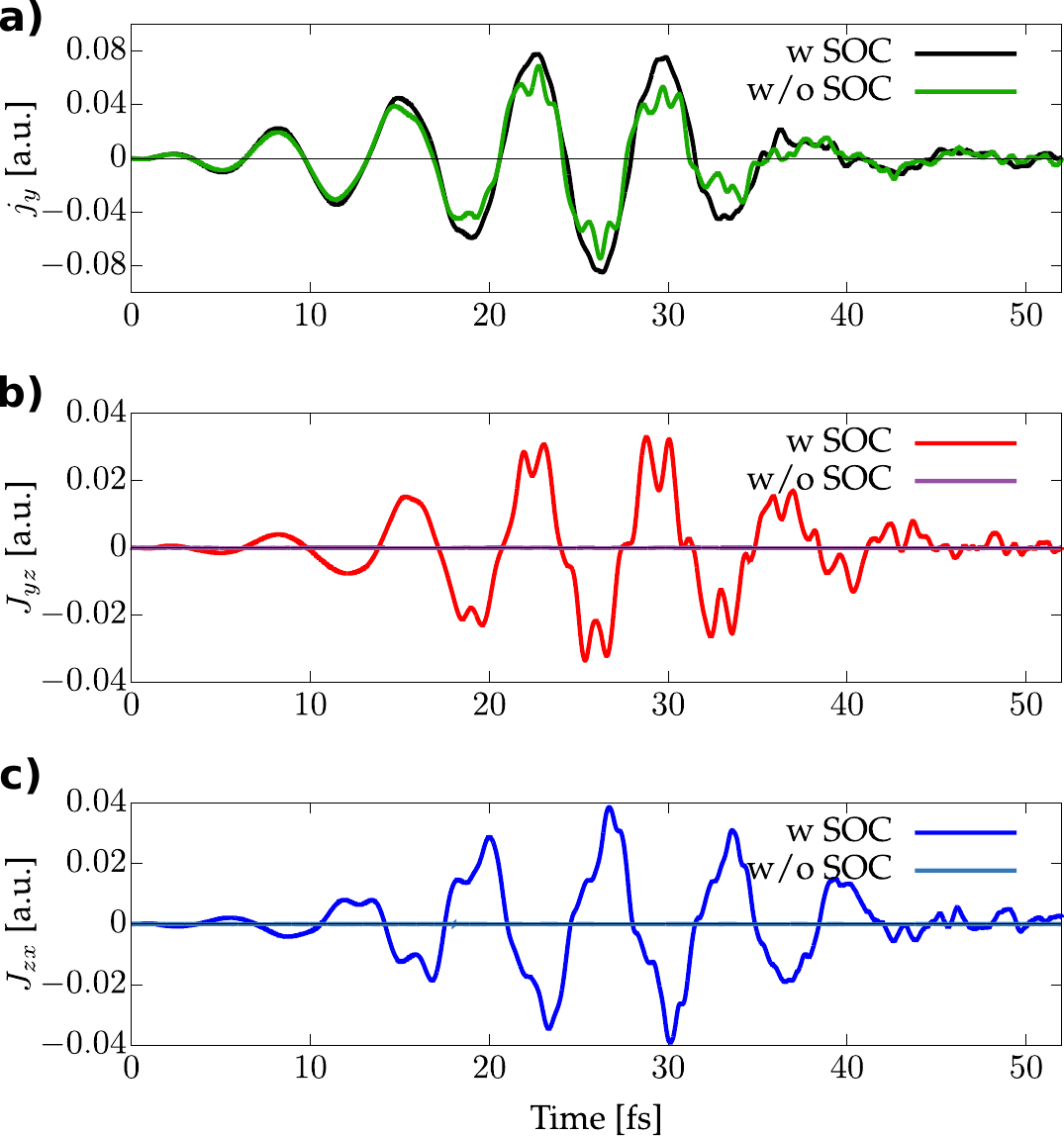}
    \caption{\label{fig:currents} a) Calculated charge current along the $y$ axis, for a RHS circularly-polarized driver in the $y-z$ plane with and without SOC. b) and c) show respectively the $J_{yz}$ and $J_{zx}$ components of the corresponding spin current. See the main text for more details.}
  \end{center}
\end{figure}

\begin{figure}[ht]
  \begin{center}
    \includegraphics[width=0.9\columnwidth]{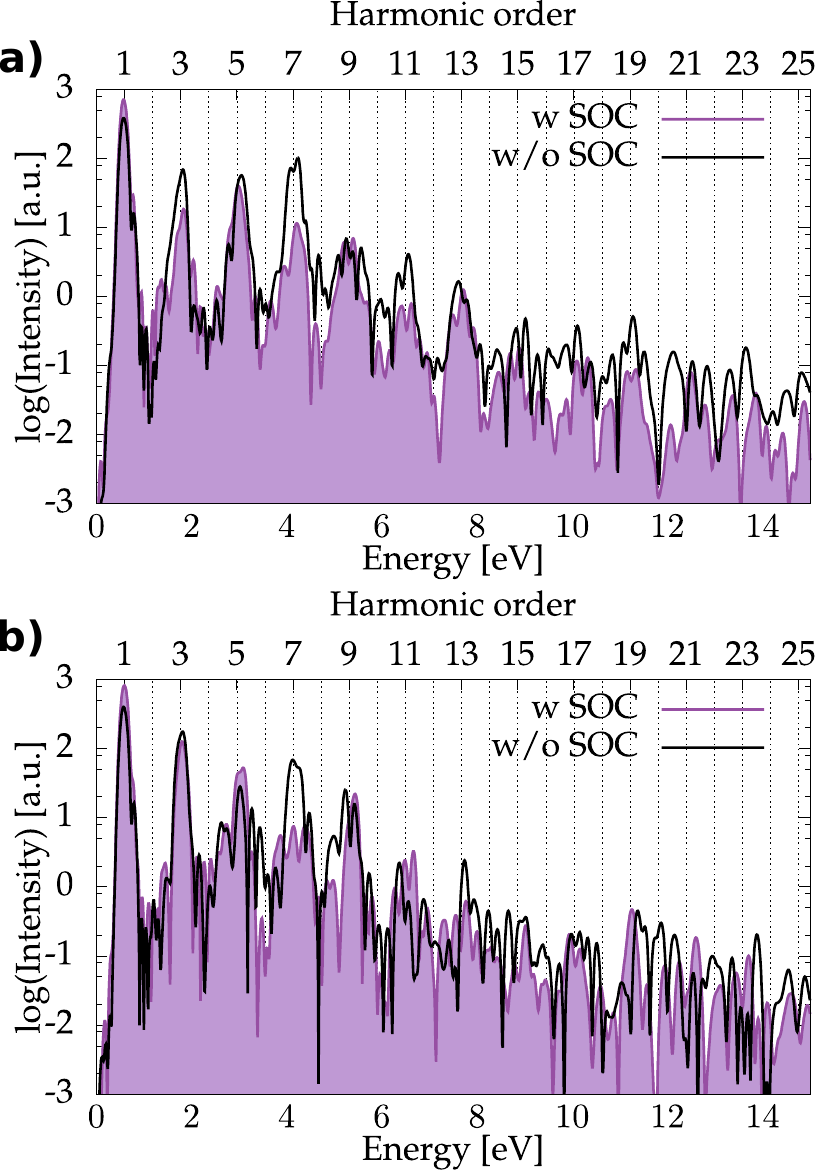}
    \caption{\label{fig:currents_HHG} Calculated HHG spectra with (red curves) and without SOC (blue curves) for a) a laser RHS circularly polarized in the $y-z$ plane, and b) a laser linearly polarized along the $y$ axis. }
  \end{center}
\end{figure}

\begin{figure}[ht]
  \begin{center}
    \includegraphics[width=0.9\columnwidth]{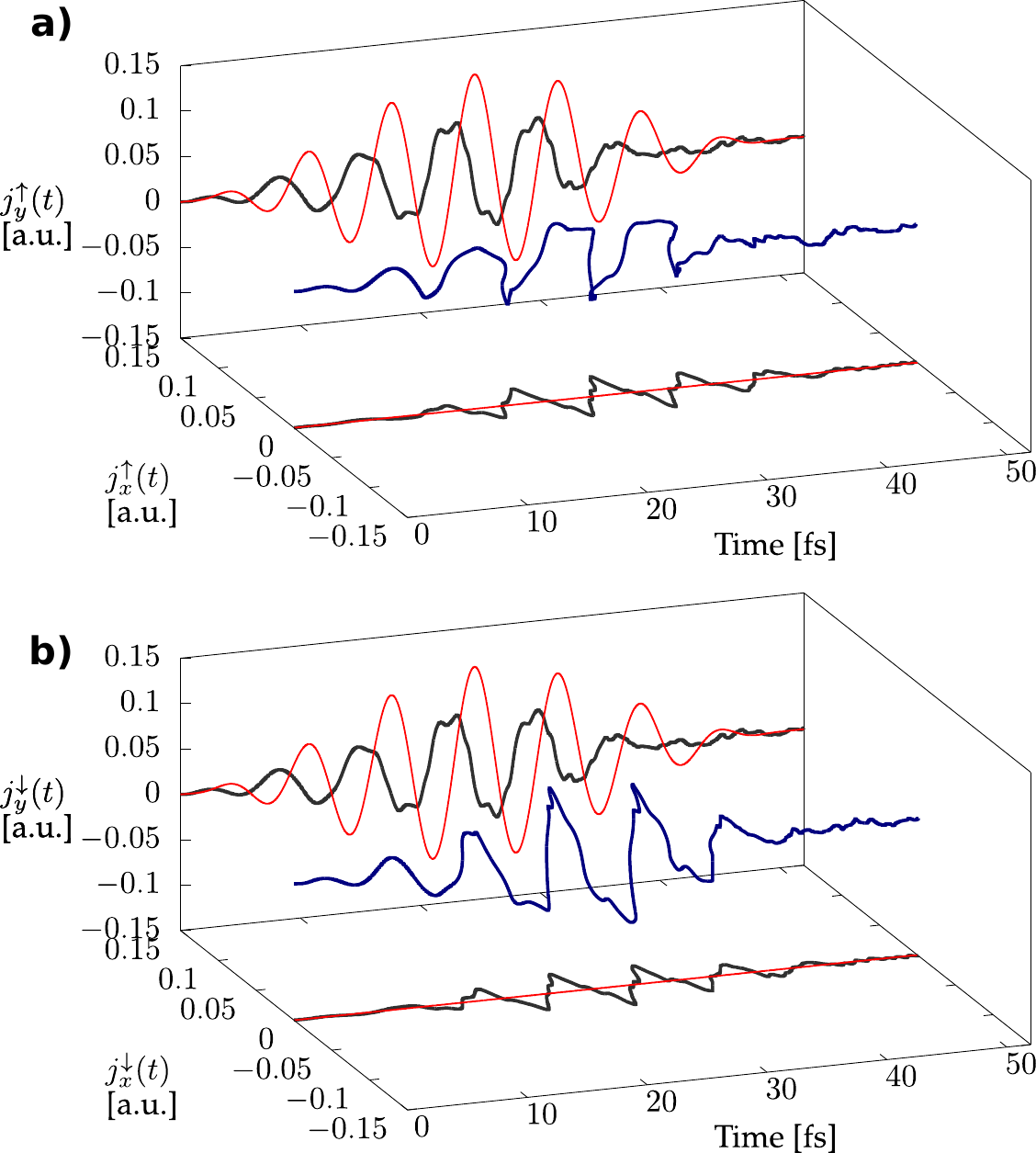}
    \caption{\label{fig:charge_current} Calculated electronic current for the a) up spin channel, and b) down spin channel for a linearly-polarized laser field along the $y$ axis. The time profile of the vector potential is shown in red.  }
  \end{center}
\end{figure}

\begin{figure}[ht]
  \begin{center}
    \includegraphics[width=0.9\columnwidth]{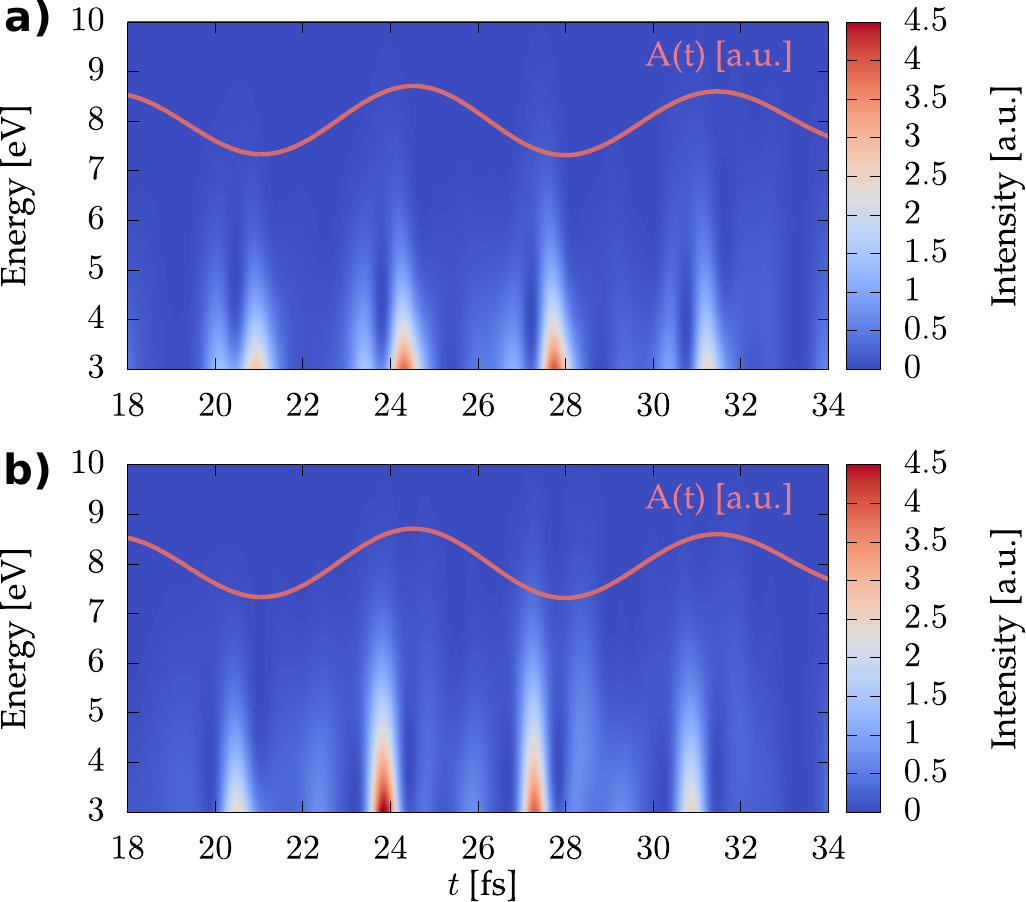}
    \caption{\label{fig:gabor} Time-frequency analysis of the harmonic emission driven by a linearly-polarized laser field, a) with SOC, and b) without SOC. The width of the Gabor transform is taken as a third of the period of the laser pulse. The time profile of the driving laser field is also shown as a solid line. }
  \end{center}
\end{figure}

Spin-orbit coupling is a key physical ingredient needed to properly describe the topological nature of Na$_3$Bi. It is therefore interesting to investigate how much the  strong-field electron dynamics is affected by the present of SOC.
As discussed in the introduction, the spin-orbit coupling introduces a spin-dependent gauge vector field $\mathcal{A} = \frac{\hbar}{4mc}\vec{\sigma}\times\vec{\nabla}V$. 
This leads to a different velocity of the electrons, depending on their spin. This directly implies that alongside with a charge current, there will be spin currents generated by the intense driving field when SOC is included.\\
We therefore computed both the charge current and the spin-current induced by the laser field, defined in second quantization as
\begin{equation}
 \hat{J}_{ij}(\vec{x}) = -\frac{i\hbar}{2m} \Big[ \hat{\psi}^\dag(\vec{x})\hat{\sigma}_i (\nabla_j\hat{\psi}(\vec{x})) - (\nabla_j\hat{\psi}(\vec{x}))^\dag \hat{\sigma}_i\hat{\psi}(\vec{x}) \Big]\,,
\end{equation}
where $\hat{\sigma}_i$ are the Pauli matrices.\\
The first question that one could rise is how important are these spin currents, compared to the magnitude of the charge current, especially for a nonmagnetic material like Na$_3$Bi.
We found that the magnitude of the non-zero components of the macroscopic spin currents ($J_{xx}$, $J_{zx}$, $J_{yy}$, and $J_{yz}$ for a laser circularly polarized in the $y-z$ plane, $J_{zx}$, $J_{zy}$, $J_{xz}$, $J_{yz}$ for a laser circularly polarized in the $x-y$ plane) all have very similar magnitude, and we found that this magnitude is only half of the magnitude of the charge current, as shown in Fig.~\ref{fig:currents} for the case of a circularly-polarized driver.
As expected, we obtain that without spin-orbit coupling, no spin currents are induced in the material.
This shows that strong driving fields can generate very efficiently spin currents, even when the electrons with different spin ``feel'' the same bandstructure. We found (not shown here) that the harmonic spectra of the different components of the spin currents display an odd-harmonic structure, and exhibit the same energy cutoff as the charge current, in agreement with the results of Ref.~\cite{PhysRevB.102.081121} for Rashba and Dresselhaus models of SOC.
The direct connection between spin currents and charge current, and thus HHG, is derived analytically and discussed in Sec.~\ref{sec:EOM}, and we therefore focus in the following of this section on the effect of the SOC on harmonic response of the charge current.\\
As shown in Fig.~\ref{fig:currents}, the SOC modifies the dynamics charge current. In fact, it reduces the harmonic response of the material, as shown in the HHG spectra, see Fig.~\ref{fig:currents_HHG}. Indeed, we obtain that SOC increases the linear response of the material while decreasing the yield of the harmonics for both polarization considered here. This is counter-intuitive, as we showed before that the SOC leads to more electrons excited in the conduction bands. This directly show that we cannot simply understand the effect of SOC on the basis of a modified band structure, but that we need to take into account how SOC reshape the electrons' dynamics within these bands.
The effect of the SOC on the electron dynamics can be better understood from the individual contribution of each spin channel to the charge current ($\vec{j}(\vec{r},t) = \vec{j}^{\uparrow\uparrow}(\vec{r},t)  + \vec{j}^{\downarrow\downarrow}(\vec{r},t)$). As shown in Fig.~\ref{fig:charge_current}, the spin-dependent effective magnetic field induced by the SOC leads to a transverse motion of each individual spins, thus leading to an elliptically polarized emission from the individual spin channels, even for a linearly-polarized electric field. As the effective magnetic field flips sign for each spin channel, the current associated with each of the two spin channels are of opposite ellipticity. 
However, because spins are degenerate in Na$_3$Bi, the transverse currents associated with each spin channel cancel each other, and no net transverse current is obtained for a linearly polarized driving field.\\
The component of the spin-resolved current along the direction of the laser is also modified, as shown in Fig.~\ref{fig:currents}. However, this change cannot be easily distinguished from the change in the electronic band curvature, also induced by the SOC. 
This modification of electrons' velocity also leads to modified motion of the carriers in the bands, and we therefore expect a time-delay in the harmonic emission compare to the case in absence of SOC, due to the transverse motion of the carriers. This is indeed what we obtain from our simulations. Indeed, as clearly visible in the time-frequency analysis plot of Fig.~\ref{fig:gabor}, the main emission event occurs at a later time when SOC is included. The side structures are also shifted in a similar way.
We found (not shown here) that a similar time delay occurs for circularly polarized laser fields, confirming that this effect is present irrespective of the polarization state of the driving pulse.\\
Let us now make some connection to some recent results obtained for the topological phase transition in the Haldane model. The increase of the number of excited electrons, as well as a time delay in the emission time of the harmonics has been identified in this model case as signatures of non-trivial topological phase in the Haldane model\cite{silva2019topological}, and the origin of these changes was associated to the Berry curvature of the material. Our results show that similar effects are obtained if we include SOC, in a material which has no Berry curvature. The reason for this is that the Berry curvature can be seen as a momentum-space effective magnetic field~\cite{PhysRevLett.113.190403}, which therefore also affects carriers trajectories~\cite{RevModPhys.82.1959}. 
How electrons evolve in materials having both SOC and Berry curvature, and which effect dominates will of course require more detailed analysis, that goes beyond the scope of the present work.

%
\section{Equation of motion of the charge current}
\label{sec:EOM}
The equation of motion of the charge current, in absence of SOC, has been shown to provide valuable insights, for instance in the context of HHG in solids~\cite{PhysRevLett.118.087403}.
Therefore, in this section, we aim at deriving the equation of motion for the physical charge current, in presence of spin-orbit coupling. 
\\
We first consider the single-particle Pauli Hamiltonian describing a particle of mass $m$ and charge $e$ in an external electromagnetic field, 
\begin{eqnarray}
 \hat{h}(t) = \frac{1}{2m} \hat{\vec{\Pi}}^2(t) - eV(t) + \frac{e\hbar}{2mc}\vec{B}(t)\ldotp \hat{\boldsymbol{\sigma}} \nonumber\\
 + \frac{e\hbar}{8m^2c^2}\big[ \hat{\vec{\Pi}}(t)\ldotp(\hat{\boldsymbol{\sigma}} \times \vec{E}(t)) + (\hat{\boldsymbol{\sigma}} \times \vec{E}(t)) \ldotp\hat{\vec{\Pi}}(t) \big],
 \label{eq:pauli}
\end{eqnarray}
where the canonical momentum is given by $\hat{\vec{\Pi}}(t) = \frac{\hbar}{i}\vec{\nabla} + \frac{e}{c}\vec{A}(t)$, the magnetic field $\vec{B}(t)$ is given by $\vec{B}(t)=\vec{\nabla}\times\vec{A}(t)$, where $\vec{A}(t)$ is the vector potential of the external field, and the electric field is given by $\vec{E}(t) = -\nabla V(t) -\frac{1}{c}\partial_t \vec{A}(t)$, with $V$ the external scalar potential acting on the electrons. 
The first two terms describe respectively the kinetic energy and the potential energy of the particle. The third one is the Zeeman term, and the last one corresponds to the spin-orbit coupling term. Importantly, the SOC contained in Eq.~\ref{eq:pauli} allow for describing the part of the spin-orbit coupling induced by the light, which is responsible for important physical effects, such as the inverse Faraday effect~\cite{PhysRevB.92.100402}.\\
While the previous Hamiltonian has been considered in many theoretical work, it is possible to obtain a local $U(1)\times SU(2)$ symmetry by including a term of the higher order $O(1/m^3)$ as explained in Ref.~\onlinecite{RevModPhys.65.733}. The motivation for this choice lies in the fact that this will allow us later to define a $U(1)\times SU(2)$ gauge-invariant spin-current, which we defined in this work as the physical spin current.
For this, we add the term $ \frac{e^2\hbar^2}{32 m^3 c^4} (\hat{\boldsymbol{\sigma}} \times \vec{E}(t))\ldotp(\hat{\boldsymbol{\sigma}} \times \vec{E}(t)) = \frac{e^2\hbar^2}{16 m^3 c^4} |\vec{E}(t)|^2$, which is half of the corresponding term in the Foldy-Wouthuysen expansion~\cite{RevModPhys.65.733}. 
Doing so, one arrives to a locally $U(1)\times SU(2)$ gauge-invariant single-particle  Hamiltonian 
\begin{equation}
 \hat{h}(t) = \frac{1}{2m} \Big[\hat{\vec{\Pi}}(t) +  \frac{e\hbar}{4mc^2}(\hat{\boldsymbol{\sigma}} \times \vec{E}(t))\Big]  ^2 - eV(t) + \frac{e\hbar}{2mc}\vec{B}(t)\ldotp \hat{\boldsymbol{\sigma}} \,,
 \label{eq:hamiltonian}
\end{equation}
which is the one we consider in the following.\\
In second quantization, all operators are expressed in terms of the field operators $\hat{\psi}(\vec{x})$ and $\hat{\psi}^\dag(\vec{x})$.
The many-body Hamiltonian $\hat{H}(t)$ consists of the time-dependent one-body part $\hat{h}(t)$ and the particle-particle interaction term $w(\vec{x},\vec{x'})$, and reads as
\begin{eqnarray}
 \hat{H}(t) = \sum_{\sigma\sigma'}\int d\vec{r} \hat{\psi}^\dag(\vec{r},\sigma) h_{\sigma\sigma'}(\vec{r},t)\hat{\psi}(\vec{r},\sigma') \nonumber\\
 + \frac{1}{2}\int d\vec{x} \int d\vec{x'} w(\vec{x},\vec{x'}) \hat{\psi}^\dag(\vec{x})\hat{\psi}^\dag(\vec{x'})\hat{\psi}(\vec{x'})\hat{\psi}(\vec{x})\,.
\end{eqnarray}
We now define the operators 
\begin{eqnarray}
 \hat{t}(\vec{x}) &=& \frac{\hbar^2}{2m}(\vec{\nabla}\hat{\psi}(\vec{x}))^\dag  (\vec{\nabla}\hat{\psi}(\vec{x}))\,,\nonumber\\
 \hat{n}(\vec{x}) &=& \hat{\psi}^\dag(\vec{x}) \hat{\psi}(\vec{x}) \,, \nonumber\\
 \hat{\vec{m}}(\vec{x}) &=& \hat{\psi}^\dag(\vec{x}) \hat{\boldsymbol{\sigma}}\hat{\psi}(\vec{x})\,, \nonumber\\
 \hat{\vec{j}}(\vec{x}) &=& -\frac{i\hbar}{2m} \Big[ \hat{\psi}^\dag(\vec{x}) (\vec{\nabla}\hat{\psi}(\vec{x})) - (\vec{\nabla}\hat{\psi}(\vec{x}))^\dag \hat{\psi}(\vec{x}) \Big] \,,\nonumber\\
 \hat{\vec{J}}(\vec{x}) &=& -\frac{i\hbar}{2m} \Big[ \hat{\psi}^\dag(\vec{x})\hat{\boldsymbol{\sigma}} (\vec{\nabla}\hat{\psi}(\vec{x})) - (\vec{\nabla}\hat{\psi}(\vec{x}))^\dag \hat{\boldsymbol{\sigma}}\hat{\psi}(\vec{x}) \Big]\,,\nonumber
\end{eqnarray}
which are respectively the kinetic energy, density, magnetization, current and spin-current operators.\\
In order to derive the equation of motion of the different observable, we start by writing the equation of motion of the creation and annihilation operators in the Heisenberg picture\cite{stefanucci2013nonequilibrium}
\begin{eqnarray}
 i\hbar\frac{d}{dt} \hat{\psi}(\vec{x},t) = \sum_{\sigma'}  h_{\sigma\sigma'}(\vec{x},-i\overrightarrow{\vec{\nabla}},t)\hat{\psi}(\vec{x},\sigma',t) \nonumber\\
 + \int d \vec{x'} w(\vec{x},\vec{x'}) \hat{n}(\vec{x'}) \hat{\psi}(\vec{x},t) \,,
\end{eqnarray}
and
\begin{eqnarray}
 i\hbar\frac{d}{dt} \hat{\psi}^\dag(\vec{x},t) = -\sum_{\sigma'}  \hat{\psi}^\dag(\vec{x},\sigma',t)h_{\sigma'\sigma}(\vec{x},i\overleftarrow{\vec{\nabla}},t) \nonumber\\
 - \int d \vec{x'} w(\vec{x},\vec{x'}) \hat{\psi}^\dag(\vec{x},t)\hat{n}(\vec{x'})  \,.
\end{eqnarray}
Following Ref.~\onlinecite{stefanucci2013nonequilibrium}, we split the equation of motion of the field operators in terms of a contribution without any external potential nor field, and the part containing them.
\begin{widetext}
We obtain that
\begin{eqnarray}
 i\hbar\frac{d}{dt} \hat{\psi}(\vec{x},t) = i\hbar\frac{d}{dt} \hat{\psi}(\vec{x},t)\Big|_{V=\vec{A}=0}  
 + \frac{ie}{2mc}\sum_p\Big[ 2A_p(t)\partial_p + (\partial_pA_p(t)) \Big]\hat{\psi}(\vec{x},t)
 + u\hat{\psi}(\vec{x},t)\nonumber\\
 + \frac{e\hbar}{8m^2c^2}\big[ \hat{\vec{\Pi}}(t)\ldotp(\hat{\boldsymbol{\sigma}} \times \vec{E}(t)) + (\hat{\boldsymbol{\sigma}} \times \vec{E}(t)) \ldotp\hat{\vec{\Pi}}(t) \big] \hat{\psi}(\vec{x},t)
 + \frac{e\hbar}{2mc}\vec{B}(t)\ldotp \hat{\boldsymbol{\sigma}}\hat{\psi}(\vec{x},t)\,,
\end{eqnarray}
and
\begin{eqnarray}
 i\hbar\frac{d}{dt} \hat{\psi}^\dag(\vec{x},t) = i\hbar\frac{d}{dt} \hat{\psi}^\dag(\vec{x},t)\Big|_{V=\vec{A}=0}  
 + \frac{ie}{2mc}\sum_p\Big[ 2A_p(t)\partial_p + (\partial_pA_p(t)) \Big]\hat{\psi}^\dag(\vec{x},t)
 - u\hat{\psi}(\vec{x},t)\nonumber\\
 - \frac{e\hbar}{8m^2c^2}\big[ \hat{\vec{\Pi}}^*(t)\ldotp(\hat{\boldsymbol{\sigma}} \times \vec{E}(t)) + (\hat{\boldsymbol{\sigma}} \times \vec{E}(t)) \ldotp\hat{\vec{\Pi}}^*(t) \big] \hat{\psi}^\dag(\vec{x},t)
 - \frac{e\hbar}{2mc}\vec{B}(t)\ldotp \hat{\boldsymbol{\sigma}}\hat{\psi}^\dag(\vec{x},t)\,,
\end{eqnarray}
where we defined
\begin{equation}
 u = -eV(\vec{x},t) + \frac{e^2}{2mc^2}|\vec{A}(\vec{x},t)|^2 - \frac{e^2\hbar^2}{16m^3c^4}|\vec{E}(\vec{x},t)|^2\,.
\end{equation}
\end{widetext}
Before obtaining the equation of motion for the physical charge current, we first need to define it. Such definition can be obtained from the continuity equation, which defines the conserved charge current for a given Hamiltonian. After some algebra, we arrive to the equation of motion of the charge density, the continuity equation,  
\begin{equation}
 \frac{\partial}{\partial t} \hat{n}(\vec{x},t) = -\vec{\nabla}\ldotp (\hat{\vec{j}}(\vec{x},t) + \hat{\vec{j}}_d(\vec{x},t) + \hat{\vec{j}}_{\mathrm{SO}}(\vec{x},t)) \,.
 \label{eq:EOM_charge}
\end{equation}
where the physical, conserved, charge current in presence of spin-orbit coupling, contains the paramagnetic current ($\hat{\vec{j}}$), the diamagnetic current ($\hat{\vec{j}}_d$), and a spin-orbit current, $\hat{\vec{j}}_{\mathrm{SO}}(\vec{x},t) = \frac{e\hbar}{4m^2c^2} \hat{\vec{m}} \times \vec{E}(t)$.\\
This equation determines the physical conserved current up to a rotational part. However, as we are interested in the time-derivative of the macroscopic current (source term of the Maxwell equations), this rotational part of the current is irrelevant, and in the following, we use Eq.~\ref{eq:EOM_charge} to define the physical current associated with the Hamiltonian of Eq.~\ref{eq:hamiltonian}.\\
In order to obtain the equation of the conserved total physical current $\hat{\vec{j}}_{\mathrm{phys}} = \hat{\vec{j}} + \hat{\vec{j}}_d + \hat{\vec{j}}_{\mathrm{SO}} $, we need the equation of motion of the magnetization density and of the paramagnetic current.
Following the same approach as for the charge density, we obtain the equation of motion of the $k$ component of the magnetization density
\begin{eqnarray}
\frac{\partial}{\partial t}\hat{m}_{k}(\vec{x},t) = 
  - \sum_p \partial_p \Big[ \hat{J}_{phys,kp}  \Big]
  + \frac{e}{mc} \sum_{qr}\epsilon_{kqr} B_q \hat{m}_{r} \nonumber\\
 +\frac{e}{2mc^2}\sum_{p} \Big[
            E_p \hat{J}_{phys,pk} - E_k \hat{J}_{phys,pp}\Big]\,,
\end{eqnarray}
where, by analogy to the continuity equation for the charge, we can defined a physical spin current as
\begin{equation}
 \hat{J}_{phys,kp} = \Big[ \hat{J}_{H,kp} + \frac{e}{mc} \hat{m}_{H,k} A_p - \frac{e\hbar}{4m^2c^2}\sum_{r}\epsilon_{kpr} E_r \hat{n}_H  \Big]\,,
 \label{eq:spin_current}
\end{equation}
where we recognize a paramagnetic, a diamagnetic, and a spin-orbit spin current.  The motivation for this definition lies in the fact that the spin current defined in Eq.~\ref{eq:spin_current} is a mixed space and spin quantity, which is for our definition locally $U(1)\times SU(2)$ gauge invariant. This symmetry is the same as the one of the Hamiltonian  of Eq.~\ref{eq:hamiltonian} and motivates us to use this expression to define the physical spin current, in the same way that the physical charge current, which is a spatial quantity only, is invariant under $U(1)$ gauge transformation.
Here $\epsilon_{ijk}$ denotes the Levi-Civita symbol.\\
The equation of motion of the paramagnetic current is found to be 
\begin{eqnarray}
 \frac{\partial}{\partial t} \hat{j}_{k,H}(\vec{x},t) = -\sum_{p}\partial_p \hat{T}_{pk,H} - \hat{W}_{k,H} \nonumber\\
 - \frac{1}{m}(\partial_kw(\vec{x},t))\hat{n}_H(\vec{x},t) 
  - \frac{e\hbar}{2m^2c}\sum_p  \partial_k(B_p) \hat{m}_{p,H} \nonumber\\
 -\frac{e}{mc}\sum_p\Bigg[ \partial_p\Big(A_p\hat{j}_{k,H}
 + \frac{\hbar}{4mc} \sum_{q,r} \epsilon_{pqr}E_r \hat{J}_{H, qk} \Big) \nonumber\\
 + (\partial_k A_p)\Big[\hat{j}_{p,H} + \frac{e\hbar}{4m^2c^2}\sum_{q,r}\epsilon_{pqr}\hat{m}_{q,H} E_r \Big] \nonumber\\
 + \frac{\hbar}{4mc}\sum_{q,r} \epsilon_{pqr} (\partial_kE_r)\Big[ \hat{J}_{H, qp} 
 + \frac{e}{mc} A_p\hat{m}_{q,H}\Big]\Bigg]\,,
\end{eqnarray}
where $\hat{T}_{pk}$ is the so-called momentum-stress tensor~\cite{stefanucci2013nonequilibrium}
\begin{equation}
 \hat{T}_{pk} = \frac{1}{2m^2} \Big[ (\partial_k \hat{\psi}^\dag)(\partial_p\hat{\psi}) + (\partial_p \hat{\psi}^\dag)(\partial_k\hat{\psi}) -\frac{1}{2} \partial_k\partial_p\hat{n}\Big]\,,
\end{equation}
while the operator $\hat{W}_k$ is defined by~\cite{stefanucci2013nonequilibrium}
\begin{equation}
 \hat{W}_k(\vec{x},t) = \frac{1}{m}\int d\vec{x'} \hat{\psi}^\dag(\vec{x})\hat{\psi}^\dag(\vec{x'})(\partial_kv(\vec{x}-\vec{x'})\hat{\psi}(\vec{x'})\hat{\psi}(\vec{x}) \,.
\end{equation}
\begin{widetext}
Putting everything together, we get the following expression for the equation of motion of the physical charge current
 \begin{eqnarray}
 \frac{\partial}{\partial t} \hat{j}_{\mathrm{phys},k}(\vec{x},t) = -\sum_{p}\partial_p \Big[ \hat{T}_{pk} + \frac{e}{mc} ( A_p\hat{j}_{k,H} + \hat{j}_{phys,p} A_k  ) + \frac{e\hbar}{4m^2c^2} \sum_{qr} (\epsilon_{kqr}\hat{J}_{phys,pq}E_{r} + \epsilon_{pqr} \hat{J}_{qk}E_r )\Big]\nonumber \\
 - \hat{W}_{k} 
 -\frac{e^2\hbar}{4m^3c^3} \Bigg[(\vec{m}_H\times\vec{B})\times\vec{E}\Bigg]_k \nonumber\\
  - \frac{e}{m}\Big[ \hat{n} \vec{E}  + \frac{1}{c}\hat{\vec{j}}_{phys} \times \vec{B}\Big]_k
  +\frac{e\hbar}{4m^2c^2} \Big[\hat{\vec{m}} \times \partial_t \vec{E}\Big]_k
  - \frac{e\hbar}{2m^2c}\sum_p \hat{m}_{p}\Big( \partial_k(B_p)\Big)\nonumber\\
  + \frac{e\hbar}{4m^2c^2}\sum_{pqr} \hat{J}_{phys,pq} \Bigg[\epsilon_{kpr}\partial_q - \epsilon_{qpr} \partial_k + \epsilon_{kqr}\frac{e}{2mc^2} E_p \Bigg] ( E_{r} )\,.
  \label{eq:EOM_current}
\end{eqnarray}
It is clear that this equation of motion contains many different contributions, which are not easy to analyze.
In fact, it is easier to analyze the equation of motion of the macroscopic charge current, which this is the source term of Maxwell equation, and therefore represent the emitted electric field. In a periodic system, the interaction term $\hat{W}$, as well as the divergences (first term in Eq.~\ref{eq:EOM_current}) vanish, and therefore we obtain for the macroscopic average of the physical charge current, denoted $\bra \hat{j}_{\mathrm{phys},k}(t) \ket$, that is easier to analyze
\begin{eqnarray}
\frac{\partial}{\partial t}\bra \hat{j}_{\mathrm{phys},k}(t) \ket = \int_\Omega d\vec{x} \Bigg(
 - \frac{e}{m}\Big[ \hat{n} \vec{E}  + \frac{1}{c}\hat{\vec{j}}_{phys} \times \vec{B}\Big]_k
  +\frac{e\hbar}{4m^2c^2} \Big[\hat{\vec{m}} \times \partial_t \vec{E}\Big]_k \nonumber\\
 -\frac{e^2\hbar}{4m^3c^3} \Bigg[(\vec{m}_H\times\vec{B})\times\vec{E}\Bigg]_k
  - \frac{e\hbar}{2m^2c}\sum_p \hat{m}_{p}\Big( \partial_k(B_p)\Big)\nonumber\\
 + \frac{e\hbar}{4m^2c^2}\sum_{pqr} \hat{J}_{phys,pq} \Big(\epsilon_{kpr}\partial_qE_{r} - \epsilon_{qpr} \partial_kE_{r} + \frac{e}{2mc^2} \epsilon_{kqr}E_pE_{r} \Big) \Bigg)\,.
    \label{eq:EOM_current_macro}
\end{eqnarray}
\end{widetext}
Equation~\ref{eq:EOM_current_macro} is nothing but a force-balance equation, and the right-hand-side terms are the external forces acting of the electrons. We directly identify the first term as the Lorentz force. In absence of spin-orbit coupling, this is the only force term present in the equation of motion of the macroscopic charge current. The third term appears only when an electric field and a magnetic field coexist (see Ref.~\cite{PhysRevLett.95.187203} and references therein).
The fourth term is a spin force that appears in presence of a non-uniform magnetic field. This term was first evidence by the Stern and Gerlach experiment. The second term, which relates the magnetization of the electron to the external electric field, is also related to non-uniform magnetic field, if we assume that the external field fulfill Maxwell equations. 
Finally, we also get a spin transverse force given by the term $\Big[\vec{E}\circ\bra\hat{\vec{J}}_{phys}\ket\Big]\times \vec{E} $ (here ``$\circ$'' means a dot product on the spin index), which was discussed in Ref.~\cite{PhysRevLett.95.187203}. This force is proportional to the square of the electric field, and the spin current whose polarization is projected along the electric field. This force is the counterpart of the force acting on a charged particle in a magnetic field $\vec{j}\times\vec{B}$, in which the spin current replaces the charge current, and the electric field replaces the magnetic field. \\
%
In order to proceed with the analysis of the equation of motion, we split the electric-field $\vec{E}$ into a contribution from the electron-ion potential $v_{\mathrm{ext}}$, which we assume to be time independent (ions are clamped), and a part originating from the laser field $\vec{E}_l$, assumed to be uniform (dipole approximation). Moreover, we assume that no external magnetic field is present. This leads to three contributions:\\
i) a part coming from the laser field 
\begin{eqnarray}
\frac{\partial}{\partial t}\bra \hat{\vec{j}}_{\mathrm{laser}}(t) \ket = 
  - \frac{e}{m} N_e\vec{E}_l 
  +\frac{e\hbar}{4m^2c^2} \bra\hat{\vec{m}}\ket \times \partial_t \vec{E}_l \nonumber\\
 +\frac{e^2\hbar}{8m^3c^4} \Big[\vec{E}_l\circ\bra\hat{\vec{J}}_{phys}\ket\Big]\times \vec{E}_{l}\,,
\end{eqnarray}
ii) an external part, from the electron-ion potential 
\begin{eqnarray}
\frac{\partial}{\partial t}\bra \hat{j}_{\mathrm{ext},k}(t) \ket = -\int_\Omega d\vec{x} \Bigg(
  \frac{e}{m}\hat{n}  \Big[ \vec{\nabla} v_{\mathrm{ext}} \Big]_k\nonumber \\ 
  -\frac{e\hbar}{4m^2c^2} \sum_{pqs}\hat{J}_{phys,pq} \Big[  ( \epsilon_{kps}\partial_q \partial_s v_{\mathrm{ext}})  - (\epsilon_{qps} \partial_k \partial_s v_{\mathrm{ext}})\nonumber \\ 
 +\frac{e}{2mc^2} \epsilon_{kps}(\partial_q v_{\mathrm{ext}} ) (\partial_s v_{\mathrm{ext}}) \Big] \Bigg)\,,
 \label{eq:external}
\end{eqnarray}
and iii) a part containing cross terms coupling the electron-ion potential to the laser field
\begin{eqnarray}
 \frac{\partial}{\partial t}\bra \hat{\vec{j}}_{\mathrm{cross}}(t) \ket = \frac{e^2\hbar}{8m^3c^4}\int_\Omega d\vec{x} \Bigg(
  \Big[\vec{E}_l\circ\hat{\vec{J}}_{phys}\Big]\times (\vec{\nabla} v_{\mathrm{ext}})  \nonumber\\
  +\Big[(\vec{\nabla} v_{\mathrm{ext}})  \circ \hat{\vec{J}}_{phys} \Big]\times \vec{E}_l \Bigg)\,.
\end{eqnarray}
Let us analyze the terms we are getting. The first term of the laser contribution is the remaining of the Lorentz force. It is present without spin-orbit coupling but is not responsible for any harmonic, as it only contains the frequency components of the external laser field.
The second term is related to the macroscopic magnetization of the system. In a non-magnetic material, such as Na$_3$Bi, this term vanishes. Finally, we get an high-order term coming from the spin transverse force. We implemented this term, as it is not taken care by the pseudopotentials as this is the case for the usual SOC, and checked that it is indeed numerically negligible.
The cross terms contain high-order contribution ($O(1/c^4)$), and are therefore negligible here.
Finally, we turn our attention to the external contributions. As a first term, we obtain the part coming from the Lorentz force, plus a high-order derivative of the electron-ion potential, that can be interpreted as a mass renormalization term. Given that this is also a high-order term, we also neglect it.
In fact, among all the SOC-related s, only the second part of Eq.~\ref{eq:external} contains a term that does not scale as $1/c^4$ but as $1/c^2$ for non-magnetic materials. This term is therefore expected to be the dominant contribution when SOC is included.
\begin{widetext}
This leads to the lowest order in $1/c$ and for a non-magnetic materials such as Na$_3$Bi, to a modified expression for HHG in presence of SOC
\begin{eqnarray}
\mathrm{HHG}(\omega) \propto \sum_k\left|\mathrm{FT}\Bigg\{\int_{\Omega} d^3\mathbf{r}\, n(\mathbf{r},t) \partial_k v_{\mathrm{ext}}(\mathbf{r}) 
+\frac{\hbar}{4mc^2} \sum_{pqs}\hat{J}_{phys,pq}(\vec{r},t) \Big[ \epsilon_{kps}\partial_q \partial_s v_{\mathrm{ext}}(\vec{r})  - \epsilon_{qps} \partial_k \partial_s v_{\mathrm{ext}}(\vec{r}) \Big]
\Bigg\} + N_e\mathbf{E}(\omega)
 \right|^2\,.
\label{eq:new_formula_HHG}
\end{eqnarray}
\end{widetext}
This expression is the main result of this section. It provides important physical insights into the effect of the spin-orbit coupling on HHG in non-magnetic materials.
Indeed, this shows that the spin current is a source term for the equation of motion of the charge current, and therefore directly imprints into the emitted harmonic light. More precisely, we note that only the microscopic fluctuations of the spin current contribute to this new term, and hence the HHG cannot be directly used as a probe of macroscopic spin currents. This formula offers an alternative perspective to the effect of the SOC on the strong-field response of bulk materials. The SOC does not simply modifies the HHG by affecting the bandstructure of the materials, it also couples the charge current dynamics to the one of the spin current. This is the direct consequence of the change in velocity of the electrons, induced by the effective SU(2) gauge vector potential. 
Finally, let us comment on the fact that we considered here a non-magnetic material. If the material is magnetic, our equation of motion Eq.~\ref{eq:EOM_current_macro} contains another contribution to the lowest order in the relativistic corrections, which deserves to be investigated, in particular in presence of demagnetization. Some of the higher-order contributions are also leading to harmonic emission along different directions, where there should not be any emission if SOC is emitted and terms containing an electric field will lead to even-order harmonic emission. Therefore one might be able to measure these terms independently of the rest of the systems' response. A similar idea was already used to measure an effective Berry curvature in quartz from the measured HHG spectra~\cite{luu2018measurement}, and our equation of motion offers many directions to use a similar logic to probe microscopic details of spin and charge current. We plan to investigate the precise role of these different terms in future works.

%
\section{Conclusion}
\label{sec:conclusions}
%
In conclusion, we investigated the effect of the SOC on the strong-field electron dynamics in the topological Dirac semimetal Na$_3$Bi.
We showed that the SOC affects the strong-field response by modifying the electronic band-structure of the material, which controls the injection of carriers to the conduction bands.
Beside, we showed that the SOC modifies the velocity of the electrons in the bands by acting as a spin-dependent effective magnetic field.
As a consequence, the spin-resolved electronic current becomes elliptically polarized, even for a linearly-polarized pump laser field.
This modified trajectories modify the motion of the carriers, and lead to a time delay in the harmonic emission.
We then derived the equation of motion of the total charge current from a locally gauge-invariant U(1)$\times$SU(2) Hamiltonian properly describing the spin-orbit coupling in presence of a light field.
From this, we showed that the SOC couples the charge current to the spin current, and we derived a novel formula for the HHG, including to the lowest order relativistic corrections.
This relation between charge and spin currents opens interesting perspectives for the optical spectroscopy of spin currents.
Moreover, our equation of motion Eq.~\ref{eq:EOM_current_macro} can be applied to other materials, such as magnetic materials, for which other relativistic corrections might play a dominant role.

\acknowledgments
This work was supported by the European Research Council (ERC-2015-AdG694097), the Cluster of Excellence ‘Advanced Imaging of Matter' (AIM), Grupos Consolidados (IT1249-19) and SFB925. The Flatiron Institute is a division of the Simons Foundation.

%
%
%
%

%

\end{document}